\documentclass[aps,prd,twocolumn,showpacs,showkeys,preprintnumbers,amsmath,amssymb,amsfonts]{revtex4}
\usepackage{bm}

\newcommand{\refb}[1]{(\ref{#1})}
\def\dd{{\mathrm{d}}}

\begin{document}

\author{M.~Cvitan}\email{mcvitan@phy.hr}
\author{S.~Pallua}\email{pallua@phy.hr}
\affiliation{
Department of Theoretical Physics,\\
Faculty of Natural Sciences and Mathematics,\\
University of Zagreb,\\
Bijeni\v{c}ka c.~32, pp.~331, 10000 Zagreb, Croatia}
\author{P.~Prester
\footnote{On leave of absence from Department of Theoretical Physics, 
Faculty of Natural Sciences and Mathematics, Zagreb, Croatia}}\email{pprester@phy.hr}
\affiliation{
Max-Planck-Institut f\"{u}r Gravitationsphysik\\
(Albert-Einstein-Institut), Am M\"{u}hlenberg 1,\\
D-14476 Golm (b.~Potsdam), Germany}
\date{\today}
\preprint{Preprint ZTF 04-03}
\title{Conformal entropy as a consequence of the properties of stationary Killing horizons}

\begin{abstract}
We show that microscopic black hole entropy formula based on Virasoro algebra can be derived from usual properties of stationary Killing horizons alone and absence of singularities of curvature invariants on them. In such a way some usual additional assumptions are shown to be fulfilled. In addition, for all quantities power expansion near  horizon and thus explicit insight of the limiting procedure is given. More important the near horizon conformal symmetry proposed by Carlip together with its consequences on microscopic entropy is given a clear geometric origin.
\end{abstract}
\pacs{04.70.Dy, 11.25.Hf, 04.60.-m, 04.50.+h}
\keywords{black holes; entropy; conformal symmetry}

\maketitle

\section{Introduction}
One of the promising efforts to understand microscopic origin of black hole
entropy is due to Solodukhin \cite{Solodukhin:1998tc} and Carlip \cite{Carlip:2002be,Carlip:2001kk,Carlip:1998wz,Carlip:1999cy} approach which both try to exploit
conformal symmetry and corresponding Virasoro algebra. In particular, Carlip
in the case of Einstein gravity assumes a certain class of boundary
conditions near horizon which enable one to identify a subalgebra of algebra of
diffeomorphisms which turns out to be Virasoro algebra. Calculations of
central charge then enabled to calculate entropy via Cardy formula. The refinements and open questions of this method have been discussed in various references \cite{Carlip:1999cy,ParkCarlip,Dreyer:2001py,Silva:2002jq,Jing:2000yd}. 
These results have been subsequently generalised to Gauss--Bonnet
gravity and for higher curvature Lagrangians \cite{Cvitan:2002cs,Cvitan:2002rh,Cvitan:2003vq}. This is the clear indication
that these properties are properties of horizons and depend only on
diffeomorphism invariance of gravity but not much of particular form of
interaction. 

However, there have been several open questions in this method which have to be answered. For instance it is important to construct
examples where the boundary conditions imposed by Carlip or their
consequences are indeed realised. Also the derivation required additional assumptions. This refers in particular to the assumptions on behaviour of the so called ``spatial derivatives'' (assumed in Appendix~A of ref.~\cite{Carlip:1999cy}).
In the subsequent
generalisations \cite{Cvitan:2002rh,Cvitan:2003vq} to higher order interactions this was even more important.

Also, from the conceptual point of view it would be desirable to add more
understanding of the origin of the obtained
properties. 
In fact one suggestion for its physical interpretation is given in the framework of induced gravity \cite{Frolov:2003ed}. The other proposal is that these properties have geometrical origin. Indeed, recently in \cite{Medved:2004tp} it was shown that existence of
stationary Killing horizon together with absence of curvature singularities on horizon implies very restricted geometry near the horizon
and also leads to conformal properties of Einstein tensor. It was also suggested
that these properties could be the realisation of Carlip boundary
conditions. 

In this paper we assume existence of such stationary black hole
horizon. We want to show that the boundary conditions proposed in
\cite{Carlip:1999cy} are then indeed realised. In addition we want to show that
properties of stationary horizons enable to calculate the necessary
quantities for central charge and entropy. It will be possible to do the
explicit calculation to leading order but also to next orders which vanish
when we perform integration over horizon. In such a way the conformal
symmetry and Virasoro algebra have indeed geometrical interpretation in the sense that they are consequence of horizon properties, as suggested in \cite{Medved:2004tp}.

\section{Boundary properties at Killing horizons}

Axially symmetric black holes have two Killing vectors e.g.
\begin{equation}\label{e1}
t^a = \left(\frac{\partial}{\partial t}\right)^a
\thickspace\textrm{,}\qquad
\phi^a = \left(\frac{\partial}{\partial \phi}\right)^a
\thickspace\textrm{,}
\end{equation} 
with corresponding coordinates $t$, $\phi$.
The other two coordintes $n$, $z$ can be chosen so that in the equal time hypersurface one chooses Gauss normal coordinate $n$ ($n=0$ on the horizon) and remaining coordinate $z$ so that metric has the form 
\begin{eqnarray}\label{e2}
\dd s^2 & = & - N(n,z)^2 \dd t^2 + g_{\phi\phi}(n,z) 
\left( \dd\phi - \omega(n,z) \dd t \right)^2 
\\\nonumber 
&+& \dd n^2+ g_{zz}(n,z) \dd z^2
\thickspace\textrm{.}
\end{eqnarray} 
Horizon is defined with
\begin{equation}\label{e3}
N(n,z)=0
\thickspace\textrm{.}
\end{equation} 
Now, well known theorems imply
\begin{equation}\label{e4}
\kappa \equiv \lim_{n\to 0}\partial_{n}N 
= \textrm{const \ensuremath{> 0}}
\thickspace\textrm{,}
\end{equation} 
on horizon (for nonextremal black holes), also
\begin{equation}\label{e5}
\Omega_H \equiv
-\lim_{n\to 0} \frac{g_{\phi t}}{g_{\phi\phi}} =
\lim_{n\to 0} \omega = \textrm{constant on horizon}
\thickspace\textrm{,}
\end{equation} 
and the property that horizon is extrinsically flat with the consequence
\begin{equation}\label{e6}
\lim_{n\to 0} \frac{\partial g_{\mu\nu}}{\partial n} = 0
\quad\textrm{on horizon}
\thickspace\textrm{.}
\end{equation} 
Absence of curvature singularities on horizon implies \cite{Medved:2004tp} that metric coefficients have following Taylor expansions:
\begin{eqnarray}\label{e7}
N(n,z)&=& \kappa n + \frac{1}{3!} \kappa_2(z) n^3 +  O(n^4)\\\nonumber
g_{\phi\phi}(n,z) &=&  g_{H\phi\phi}(z) + 
\frac{1}{2} g_{2\phi\phi}(z) n^2 + O(n^3)\\\nonumber
g_{zz}(n,z) &=&  g_{Hzz}(z) + 
\frac{1}{2} g_{2zz}(z) n^2 + O(n^3)\\\nonumber
\omega(n,z) &=& \Omega_H + \frac{1}{2} \omega_2(z) n^2 +  O(n^3)
\thickspace\textrm{.}
\end{eqnarray} 
The Killing horizon has a Killing null vector
\begin{equation}\label{e8}
\chi^a = t^a + \Omega_H \phi^a
\thickspace\textrm{.}
\end{equation} 
On horizon ($\chi^2=0$), this vector satisfies well known relation
\begin{equation}\label{e9}
\nabla^a \chi^2 = -2\kappa \chi^a
\thickspace\textrm{.}
\end{equation} 
For $\chi^2 \geq 0$, left hand side of \refb{e9} defines vector $\rho_a$
\begin{equation}\label{e11}
\nabla_a \chi^2 = -2\kappa \rho_a
\thickspace\textrm{.}
\end{equation} 

In the following we shall use the basis
\begin{equation}\label{e13}
\chi^a\thickspace\textrm{,}\quad
\phi^a\thickspace\textrm{,}\quad
\rho^a\thickspace\textrm{,}\quad
z^a\thickspace\textrm{.}
\end{equation} 
Explicitly
\begin{eqnarray}\label{e14}
&&\chi^a
= \begin{pmatrix}
1 \\ 
\Omega_H \\ 
0 \\ 
0
\end{pmatrix}\thickspace\textrm{,}\quad
\phi^a
= \begin{pmatrix}
0 \\ 
1 \\ 
0 \\ 
0
\end{pmatrix}\thickspace\textrm{,}\quad
\\\nonumber
&&\rho^a
= \begin{pmatrix}
0 \\ 
0 \\ 
\kappa n + O(n^3)\\ 
O(n^4)
\end{pmatrix}\thickspace\textrm{,}\quad
z^a
= \begin{pmatrix}
0 \\ 
0 \\ 
0 \\ 
1
\end{pmatrix}\thickspace\textrm{.}
\end{eqnarray} 
Leading terms of nonvanishing products of basis vectors are
\begin{eqnarray}\label{e15}
\chi\cdot\chi &=& -\kappa^2 n^2 + O(n^4)\\\nonumber
\chi\cdot\phi &=& -\frac{1}{2} {g_{H\phi \phi }(z)}\,
\omega_2(z)\,n^2 
+ O(n^3)
\\\nonumber
\phi\cdot\phi &=& g_{H\phi\phi}(z) + O(n^2)\\\nonumber
\rho\cdot\rho &=& \kappa^2 n^2 + O(n^4)\\\nonumber
\rho\cdot z &=& O(n^4)\\\nonumber
z\cdot z &=& g_{Hzz}(z) + O(n^2)
\thickspace\textrm{,}
\end{eqnarray} 
and all other products are zero
\begin{equation}
\chi\cdot\rho =
\chi\cdot z =
\phi\cdot\rho =
\phi\cdot z = 0
\thickspace\textrm{.}
\end{equation} Now we can, following Carlip, consider diffeomorphisms generated by following vector fields
\begin{equation}\label{e16}
\xi^a=T\chi^a+R\rho^a
\thickspace\textrm{.}
\end{equation} 
In principle diffeomorphisms could change the position of
\begin{equation}\label{e16b}
\chi^2 = 0
\thickspace\textrm{.}
\end{equation} 
One requires therefore the condition that surface variations keep this surface fixed or
\begin{equation}\label{e17}
\delta \chi^2 = 0
\thickspace\textrm{,}
\end{equation} 
and even a stronger condition
\begin{equation}\label{e17b}
\frac{\delta \chi^2}{\chi^2} = 0
\thickspace\textrm{.}
\end{equation} 
Straightforward calculation shows
\begin{eqnarray}\label{e18}
\delta \chi^2 &=& \chi^a\chi^b\delta g_{ab} = \chi^a\chi^b(\nabla_a\xi_b+\nabla_b\xi_a)\\\nonumber
 &=&
2R\chi^b\nabla_\chi\rho_b+2(\nabla_\chi T)\chi^2
\thickspace\textrm{.}
\end{eqnarray}
Due to exact relation (\cite{Carlip:1999cy} eqn.~A.4)
\begin{equation}\label{e19}
\chi^b\nabla_\chi\rho_b = -\kappa \rho^2
\thickspace\textrm{,}
\end{equation} 
it follows that
\begin{equation}\label{e20}
\delta \chi^2 = -2\kappa R\rho^2 + 2(\nabla_\chi T)\chi^2 \thickspace\textrm{.}
\end{equation} 
Thus requirement \refb{e17} will be satisfied automatically on horizon, 
and the stronger requirement \refb{e17b} will be fulfilled if
\begin{equation}\label{e21}
R = -\frac{1}{\kappa}\frac{\chi^2}{\rho^2}\nabla_\chi T
\thickspace\textrm{.}
\end{equation} 

Selecting one parameter group of diffeomorphisms one can calculate commutator of two vector fields and provided we impose the additional condition on the diffeomoprhism defining functions
\begin{equation}\label{e22}
\rho^a\nabla_a T = 0\qquad\textrm{at horizon}
\thickspace\textrm{,}
\end{equation} 
one obtains closed algebra
\begin{eqnarray}\label{e23}
\{ \xi_1,\xi_2  \}^a &=& (T_1{\nabla_\chi}T_2 - T_2{\nabla_\chi}T_1)\chi^a \\\nonumber
&+&
  \frac{1}{\kappa}\frac{\chi^2}{\rho^2}{\nabla_\chi}(T_1{\nabla_\chi}T_2 - T_2{\nabla_\chi}T_1)\rho^a
\thickspace\textrm{.}
\end{eqnarray} 
As pointed in \cite{Carlip:1999cy}, this is isomorphic to $\mathit{Diff}S^1$ or $\mathit{Diff}{\bf R}$.
An additional natural requirement on fluctuations and thus on diffeomorphisms is made as usual
\begin{equation}\label{e23b}
\delta\int_{\partial C}{\hat{\bm{\epsilon}}}
   \left({\tilde\kappa} - \frac{\rho}{|\chi|}\kappa\right) = 0 
\thickspace\textrm{,}
\end{equation} 
where ${\tilde\kappa}^2 = -a^2 / \chi^2$, and $a^a = \chi^b\nabla_b\chi^a$ is the acceleration of an orbit of $\chi^a$.
As explained elsewhere it provides us with orthogonality relations for one parameter group of diffeomorphisms.

We are now in position to calculate the near horizon expansion for fluctuations $\delta g_{ab}$ using decompositions \refb{e7}. The leading terms and next to leading order are:
\begin{widetext}
\begin{eqnarray}\label{e26}
\delta g_{ab}&=&
2\chi_{\left(a\right.}\rho_{\left.b\right)}
\left(
 \frac{\ddot{T}}{{\kappa }^3{}n^2} + 
 \left(
  \frac{3{\omega_{2}}^2{}g_{H\phi\phi}\ddot{T}}{4{}{\kappa }^5} - 
  \frac{4\kappa_2{}\ddot{T}}{3{\kappa }^4} +
  \frac{\omega_{2}{\partial_\phi \dot{T}}}{2{}{\kappa }^3} 
 \right)  + 
 O(n)
\right)\\\nonumber
&+&{{\rho }_a}{{\rho }_b}
\left(
 \frac{-2{}{\dot{T}}}{{\kappa }^2{}n^2} +
 \left(
  \frac{-7{}{\omega_{2}}^2{}g_{H\phi\phi}{}{\dot{T}}}{2{}{\kappa }^4} +
  \frac{14{}\kappa_2{}{\dot{T}}}{3{}{\kappa }^3}
 \right) + 
 O(n)
\right)\\\nonumber
&+&2{{\chi }_{\left(a\right.}}{{\phi }_{\left.b\right)}}
\left(
  \left(
   \frac{-\omega_{2}{}{\dot{T}}}{2{}{\kappa }^2} +
   \frac{{\partial_\phi T}}{g_{H\phi\phi}}
  \right) -
  \frac{2{}{}\omega_{3}{}{\dot{T}}}{{\kappa }^2}{}n +
  O(n^2)
\right)\\\nonumber
&+&2{{\rho }_{\left(a\right.}}{{\phi }_{\left.b\right)}}
\left(
 \left( 
  \frac{\omega_{2}{}{\ddot{T}}}{2{}{\kappa }^3} -
  \frac{{\partial_\phi \dot{T}}}{g_{H\phi\phi}{}\kappa }
 \right)  + 
 \frac{\omega_{3}{}{\ddot{T}}}{{\kappa }^3}{}n +
 O(n^2)
\right)\\\nonumber
&+&{{\chi }_a}{{\chi }_b}
\left(
 \left(
  \frac{{\omega_{2}}^2{}g_{H\phi\phi}{}{\dot{T}}}{2{}{\kappa }^4} -
  \frac{\omega_{2}{}{\partial_\phi T}}{{\kappa }^2} 
 \right) +
 O(n)\right)\\\nonumber
&+&{{\phi }_a}{{\phi }_b}
\left(
 \left(
  -\frac{g_{2H\phi\phi}{}{\dot{T}}}{{g_{H\phi\phi}}^2} -
  \frac{{\omega_{2}}^2{}{\dot{T}}}{2{}{\kappa }^2}
 \right) {}n^2 + 
 O(n^3)
\right)\\\nonumber
&+&{z_a}{z_b}
\left(
 -\frac{g_{2Hzz}{}{\dot{T}}}{{g_{Hzz}}^2}{}n^2  +
 O(n^3)
\right)\\\nonumber
&+& \textrm{terms of order \ensuremath{\geqq 3}}
\thickspace\textrm{,}
\end{eqnarray}
\end{widetext}
where the expansion of metric up to the $n^4$ terms was used ($\omega_3$ is defined as $\omega(n,z) = \Omega_H + \frac{1}{2} \omega_2(z) n^2 +  \omega_3 n^3 + \ldots$).

Taking into consideration the equations \refb{e15} we can ascribe to basis vectors $\chi$, $\rho$ order $n^1$ and to $\phi$, $z$ order $n^0$. Then the above expansion is the power series containing terms up to order $n^2$.
The leading terms are
\begin{equation}
\delta g_{ab} = (\chi_a\rho_b + \rho_a\chi_b)\frac{\ddot{T}}{\kappa\rho^2} - \rho_a\rho_b\frac{2\dot{T}}{\rho^2}
\thickspace\textrm{.}
\end{equation} 

Due to the fact that our manifold has boundaries it is natural to look for central extensions of this algebra. The necessary formalism of covariant phase space was explained in
\cite{Iyer:1994ys,CovPhasSpac,Lee:nz}
and exploited in 
\cite{Carlip:1999cy,Cvitan:2002rh,Cvitan:2003vq}.
For this reason we shall mention here just the main equations and refer details to above mentioned references.
For a given Lagrangian $4$-form $\mathbf{L}$ we write the variation
\begin{equation}
\delta\mathbf{L}[\phi]=\mathbf{E}[\phi]\delta\phi + \dd\mathbf{\Theta}\left[\phi,\delta\phi\right] 
\thickspace\textrm{.}
\end{equation} 
The $3$-form $\mathbf{\Theta}$ or symplectic potential is implicitly defined in above equation. To vector fields $\xi^a$ we associate vector current $3$-form
\begin{equation}
\mathbf{J}[\xi] = \mathbf{\Theta}[\phi,\mathcal{L}_\xi\phi] - \xi\cdot\mathbf{L}
\thickspace\textrm{,}
\end{equation} 
and corresponding Noether charge $2$-form 
\begin{equation}\label{e25}
\mathbf{J} = \dd\mathbf{Q}
\thickspace\textrm{,}
\end{equation} 
It was shown in \cite{Iyer:1994ys} that Hamiltonian is a pure surface term for all diffeomorphism invariant theories and
\begin{equation}
\delta H[\xi]=\int_C
(\delta\mathbf{J}[\xi]-\dd(\xi\cdot\mathbf{\Theta}[\phi,\delta\phi]))
\thickspace\textrm{,}
\end{equation} 
where $C$ is a Cauchy surface.

Integrability condition requires that a $3$-form $\mathbf{B}$ exists with the property
\begin{equation}\label{e28}
\delta\int_{\partial C}\xi\cdot\mathbf{B}=\int_{\partial C}\xi\cdot\mathbf{\Theta}
\thickspace\textrm{.}
\end{equation} 
As explained elsewhere \cite{Carlip:1999cy} one can, starting from Hamiltonian $H[\xi]$, corresponding to some diffeomorphism $\xi^a$, write algebra of its surface terms $J[\xi]$
\begin{equation}\label{e29}
{\left\lbrace J[\xi_1], J[\xi_2] \right\rbrace}^*
= J[\left\lbrace \xi_1, \xi_2 \right\rbrace] +
K[ \xi_1, \xi_2 ]
\thickspace\textrm{,}
\end{equation} 
with
\begin{equation}\label{e30}
J[\xi]=\int_{\partial C}
\mathbf{Q}[\xi]
\thickspace\textrm{.}
\end{equation} 
and the Dirac bracket
\begin{eqnarray}\nonumber
{\left\lbrace J[\xi_1], J[\xi_2] \right\rbrace}^*=
\int_{\partial C} &&\left(
     \xi_2\cdot\mathbf{\Theta}[\phi,\mathcal{L}_{\xi_1}\phi]
   - \xi_1\cdot\mathbf{\Theta}[\phi,\mathcal{L}_{\xi_2}\phi] \right.\\
  && \left.- \xi_2\cdot(\xi_1\cdot{\bf L}) \right)\label{e31}
\thickspace\textrm{.}
\end{eqnarray}
From equation \refb{e29} one can determine central charge $K[\xi_1,\xi_2]$. Now symplectic current is \cite{Iyer:1994ys}
\begin{equation}
\Theta_{pef}=2\epsilon_{apef}
(E^{abcd}\nabla_d\delta g_{bc}-\nabla_dE^{abcd}\delta g_{bc})
\thickspace\textrm{,}
\end{equation}
where
\begin{equation}\label{e33}
E^{abcd}=\frac{\partial L}{\partial{R_{abcd}}}
\thickspace\textrm{.}
\end{equation} 
This expression is valid for Lagrangians 
which do not contain derivatives of Riemann tensors.

In this paper we are interested primarily in Einstein gravity case, but we shall include also the most general Lagrangian with quadratic terms in Riemann tensor or
\begin{equation}\label{lag}
L = \frac{1}{16\pi} R +
 \alpha R^2 + \beta R_{\mu\nu} 
R^{\mu\nu} + \gamma R_{\mu\nu\rho\sigma}R^{\mu\nu\rho\sigma}
\thickspace\textrm{.}
\end{equation}

The integrals \refb{e30} and \refb{e31} are taken over $2$-dimensional surface $\mathcal{H}$ which is intersection of Killing horizon $\chi^2 = 0$ with the Cauchy surface $C$. The volume element is 
\begin{equation}\label{e32b}
\epsilon_{abcd} = {\hat\epsilon}_{cd}
\eta_{ab} + \dots
\thickspace\textrm{,}
\end{equation} 
where only the first term contributes to the integral, and binormal $\eta_{ab}$ is
\begin{equation}\label{e33b}
\eta_{ab} = 2\chi_{[b}N_{c]}=\frac{2}{|\chi{}|\rho}\rho_{[a}\chi_{b]} + s_{[a}\chi_{b]} 
\thickspace\textrm{,}
\end{equation} 
and $s^a = (0,D_1,0,D_2)$ is tangent to $\mathcal{H}$. $N^a$ is future directed null 
normal 
\begin{equation}\label{e34b}
N^a = k^a -\alpha\chi^a - s^a
\thickspace\textrm{,}
\end{equation}  
and
\begin{equation}\label{e34c}
k^a \equiv -\left( \chi^a - \rho^a |\chi| / \rho \right) / \chi^2
\thickspace\textrm{.}
\end{equation} 
To find symplectic potential $\mathbf{\Theta}$ and in particular to perform integration in \refb{e31} we need also the quantities $\nabla_d\delta g_{bc}$.
We calculated this quantity including the $O(n^2)$ terms.
Here we write the leading term
\begin{equation}
\nabla_d\delta g_{ab} 
 =  -2 \chi_d\chi_a\chi_b      {\ddot{T} \over \chi^4} + 
       2\chi_d\chi_{(a}\rho_{b)} 
             \left(   {\dddot{T} \over {\kappa\chi^2\rho^2}} +
                {{2 \kappa \dot{T}} \over \chi^4} \right) 
\thickspace\textrm{.}
\end{equation} 
In fact the expression \refb{e31} can be written more explicitly 
\begin{eqnarray}\nonumber
{\left\lbrace J[\xi_1], J[\xi_2] \right\rbrace}^*=
-\int_{\partial C} &\hat{\bm{\epsilon}}&
\left\lbrace 2\left(
  X^{(12)}_{abcd}E^{abcd}
  -\tilde{X}^{(12)}_{abc}\nabla_d E^{abcd}
\right)\right.\\
&-& \left.
\vphantom{\left(X^{(12)}_{abcd}\right)}
\xi_2^a\xi_1^b\eta_{ab}L\right\rbrace
\label{e31b}
\thickspace\textrm{.}
\end{eqnarray}
Here, 
\begin{equation}
X^{(12)}_{abcd} = \xi_{1}^{p}\eta_{ap} 
\nabla_d\delta_2g_{bc}
- (1\leftrightarrow 2)
\thickspace\textrm{,}
\end{equation} 
\begin{equation}
\tilde{X}^{(12)}_{abc} = \xi_{1}^{p}\eta_{ap}
\delta_2g_{bc}- (1\leftrightarrow 2)
\thickspace\textrm{.}
\end{equation}

It is useful to note that tensors $X^{(12)}_{abcd}$ and  $\tilde{X}^{(12)}_{abc}$ depend only on details
of black hole and its symmetry properties (diffeomorphism defining functions) but not on the form of the Lagrangian.
We have evaluated the Taylor series near horizon for $X^{(12)}_{abcd}$ and  $\tilde{X}^{(12)}_{abc}$  and Taylor series for interaction dependent tensors
$E^{abcd}$ and $\nabla_dE^{abcd}$. They allow us to establish the following properties on horizon
\begin{widetext}
\begin{equation}\label{lim1}
\lim_{n \rightarrow 0}\left(X^{(12)}_{abcd}E^{abcd}\right) = \lim_{n \rightarrow 0}\left(-\frac{1}{4}\eta_{ab}\eta_{cd}E^{abcd}
\left[(\frac{1}{\kappa}T_1\dddot{T}_2-2\kappa T_1\dot{T}_2)
- (1\leftrightarrow 2)
\right]
\right) 
\thickspace\textrm{,}
\end{equation}
\end{widetext}
\begin{equation}\label{lim2}
\lim_{n \rightarrow 0}\left(\tilde{X}^{(12)}_{abc}\nabla_d E^{abcd}\right) = 0
\thickspace\textrm{.}
\end{equation}
Of course the last term in \refb{e31b} vanishes on horizon where we expect Lagrangian \refb{lag} to be regular (as a function of curvature invariants).

It is important to realise that contrary to previous procedures we now have explicitly under control the next to leading terms in the expansion parameter $n$ (distance of horizon). We need also to calculate the Noether charge
\begin{equation}\label{eqIII41}
Q_{ef}
=
-E^{abcd}\epsilon_{abef}
\nabla_{[c}\xi_{d]}
\thickspace\textrm{,}
\end{equation} 
or 
\begin{equation}
\mathbf{Q}
=
\hat{\bm{\epsilon}}\, E^{abcd}Y_{abcd}
\thickspace\textrm{,}
\end{equation} 
where
\begin{equation}
Y_{abcd} = -\eta_{ab}\nabla_{[c}\xi_{d]}
\thickspace\textrm{.}
\end{equation} 
The tensor $Y_{abcd}$ up to terms of order 2 can also be calculated but we omit the result here.

In that case from these definitions and \refb{e29} one obtains expression for central charge
\begin{equation}
K = \int_{\mathcal{H}}
 {{\hat{\bm{\epsilon}}}\, E^{abcd}Z_{abcd}}
\thickspace\textrm{,}
\end{equation} 
\begin{equation}
Z_{abcd} = 2 X^{(21)}_{abcd} - Y_{abcd}
\thickspace\textrm{.}
\end{equation}
Now tensors $Z$ and $E$ can be explicitly calculated due to previous expansions. The leading term is then the expression for central charge obtained in previous references
\begin{eqnarray}\label{eqIII44}
K[\xi_1,\xi_2] &=&
-{1 \over 2}\int_{\mathcal{H}}\hat{\bm{\epsilon}}\, 
E^{abcd}\eta_{ab}\eta_{cd}
  {1 \over {\kappa}} (\dot{T}_1\ddot{T}_2-\ddot{T}_1\dot{T}_2)
\thickspace\textrm{.}
\end{eqnarray}
The explicit contributions in relations for $X$ and $Y$ are higher order and thus vanish at horizon $n=0$. In usual way one would then obtain expression for entropy
\begin{equation}
S = -2\pi\int_{\mathcal{H}}{\hat{\bm{\epsilon}}\, E^{abcd}\eta_{ab}\eta_{cd}}
\thickspace\textrm{.}
\end{equation}

Here we want to add a remark. Original proposal of this approach assumed a set of boundary conditions. Despite the fact that we presented a straightforward calculation based on properties of black hole it is of interest to check above mentioned assumptions. In fact using expansion \refb{e26} one can check that following conditions assumed in references \cite{Carlip:1999cy,Cvitan:2002rh,Cvitan:2003vq} are indeed valid
\begin{eqnarray}\label{bc}
&&\chi^at^a\delta g_{ab} \rightarrow 0
\thickspace\textrm{,}\qquad
\rho^a\nabla_a(g_{bc}\delta g^{bc}) = 0
\thickspace\textrm{,}\qquad\\\nonumber
&&\rho^a\nabla_a(\frac{\rho^b \delta \chi_b}{\chi^2}) = 0
\thickspace\textrm{,}\qquad
\rho^a\nabla_a(\frac{\delta \rho^2}{\chi^2}) = 0
\thickspace\textrm{.}
\end{eqnarray}
It is important to note that relations \refb{e21}, \refb{e22}, \refb{e23b} are defining the diffeomorphisms and are thus also satisfied. The explicit form of diffeomorphisms is given in previous references.

The analysis of this paper has been done for $D=4$ but there does not seem that there could be obstructions for higher dimensions. There is in fact one partial result in the case of spherical and static metric in $D$-dimensions.
\begin{equation}
\dd s^2 = -f(x) \dd t^2 + \frac{\dd x^2}{f(x)} + r^2(x)\dd \Omega_{D-1}
\thickspace\textrm{.}
\end{equation}
With explicit calculations we have checked that boundary conditions \refb{bc} and properties \refb{lim1} and \refb{lim2} are valid.

In this paper we have treated the axially symmetric black holes. As is well known axial symmetry follows from stationarity as shown by uniqueness theorems \cite{Carter} under some standard conditions. However one may be interested in situations where these conditions are not fulfilled and thus investigate horizons which are not axially symmetric. This question would require separate analysis and would presumably be technically more complicated.
\section{Conclusion}
The well known calculation of entropy via Cardy formula is based on the calculation of central charge of a subalgebra of diffeomorphism algebra on the black hole horizon. The calculations have been based on additional plausible assumptions which then led to leading terms which gave contributions on horizon and without evaluation of next to leading terms. The approach used here starts from usual properties of horizons of stationary black holes together with regularity of curvature invariants on them which then imply restrictive power series expansion for metric fluctuations near horizon \cite{Medved:2004tp}. We are then able to obtain without previously mentioned assumptions the expansions for fluctuations of the metric and its covariant derivative and consequently for the tensor $Z$ needed in the integrand of the central charge formula. The horizon limit was then possible to perform explicitly. In addition next to leading and next to next to leading terms are explicitly exhibited.

More important in such a way we have shown that near horizon geometry implies, as suggested by \cite{Medved:2004tp}, near horizon conformal symmetry formulated by Carlip with its consequences for the black hole entropy.

\begin{acknowledgments}
We would like to acknowledge the financial support under the contract
No.\ 0119261 of Ministry of Science and Technology of Republic of 
Croatia.
\end{acknowledgments}

\appendix
\section{}\label{appX}

As mentioned in the text, the important ingredient in calculations are Taylor series near horizon for various quantities like $\delta g_{ab}$, $\nabla_d\delta g_{ab}$,
$X^{(12)}_{abcd}$, $\tilde{X}^{(12)}_{abc}$, $E^{abcd}$ and $\nabla_dE^{abcd}$. In the text we have exhibited expansion for $\delta g_{ab}$ \refb{e26}. Here we present as another example expansion of tensor $X^{(12)}_{abcd}$
including terms $n^0$, $n^1$ and $n^2$  
(we also symmetrize $X^{(12)}_{abcd}$ such that $X^{(12)}_{abcd}=X^{(12)}_{cdab}$ and $X^{(12)}_{abcd}=X^{(12)}_{[ab][cd]}$, which does not change product $X^{(12)}_{abcd}E^{abcd}$):
\begin{widetext}
\begin{eqnarray}
X^{(12)}_{abcd}
&=&
{{\chi }_a}{{\rho }_b}{{\chi }_c}{{\rho }_d}
\left(
 \frac
  {\left(2{}{\kappa }^2{}{\dot{T}_2} - {\dddot{T}_2}\right){}{T_1}}
  {4{}{\kappa }^5{}n^4} + 
 C_{0101}^{-2} {}\frac{1}{n^2} +
 O(\frac{1}{n})
\right)\\\nonumber
&+&{{\chi }_a}{{\rho }_b}{{\chi }_c}{{\phi }_d}
\left(
 C_{0102}^{-2} {}\frac{1}{n^2} +
 C_{0102}^{-1} {}\frac{1}{n} +
 O(n^0)
\right)\\\nonumber
&+&{{\chi }_a}{{\rho }_b}{{\chi }_c}{z_d}
\left(
 C_{0103}^{-2} {}\frac{1}{n^2} +
 O(n^0)
\right)\\\nonumber
&+&{{\chi }_a}{{\rho }_b}{{\rho }_c}{{\phi }_d}
\left(
 C_{0112}^{-2} {}\frac{1}{n^2} +
 C_{0112}^{-1} {}\frac{1}{n} +
 O(n^0)
\right)\\\nonumber
&+&{{\chi }_a}{{\rho }_b}{{\phi }_c}{z_d}
\left(
 C_{0123}^{0} +
 O(n)
\right)%\\\nonumber
+{{\chi }_a}{{\phi }_b}{{\chi }_c}{{\phi }_d}
\left(
 C_{0202}^{0} +
 O(n)
\right)%\\\nonumber
+{{\chi }_a}{{\phi }_b}{{\chi }_c}{z_d}
\left(
 C_{0203}^{0} +
 O(n)
\right)\\\nonumber
&+&{{\chi }_a}{{\phi }_b}{{\rho }_c}{{\phi }_d}
\left(
 C_{0212}^{0} +
 O(n)
\right)%\\\nonumber
+{{\chi }_a}{{\phi }_b}{{\rho }_c}{z_d}
\left(
 C_{0213}^{0} +
 O(n)
\right)%\\\nonumber
+{{\chi }_a}{z_b}{{\chi }_c}{z_d}
\left(
 C_{0303}^{0} +
 O(n)
\right)\\\nonumber
&+&{{\chi }_a}{z_b}{{\rho }_c}{{\phi }_d}
\left(
 C_{0312}^{0} +
 O(n)\right)%\\\nonumber
+{{\chi }_a}{z_b}{{\rho }_c}{z_d}
\left(
 C_{0313}^{0} +
 O(n)
\right)%\\\nonumber
+{{\rho }_a}{{\phi }_b}{{\rho }_c}{{\phi }_d}
\left(
 C_{1212}^{0} +
 O(n)\right)\\\nonumber
&+&{{\rho }_a}{{\phi }_b}{{\rho }_c}{z_d}
\left(
 C_{1213}^{0} +
 O(n)
\right)\\\nonumber
&+& \textrm{terms related by permutation of indices according to symmetries of \ensuremath{X^{(12)}_{abcd}}}\\\nonumber
&+& \textrm{terms of order \ensuremath{\geqq 3},}\\\nonumber
&-& (1\leftrightarrow 2)
\thickspace\textrm{.}
\end{eqnarray} 
\end{widetext}
Here the coefficients of nonleading terms are lengthy algebraic expressions given in terms of diffeomorphism defining function $T_1$, $T_2$ and Taylor coefficients of metric functions and not very informative. For these reasons we do not exhibit them here.

\end{document}